\documentclass[aps,prb, twocolumn, groupedaddress, draft, showpacs, intlimits,amsmath,amssymb,floats, superscriptaddress]{revtex4}
\usepackage{bm}
\usepackage[final]{graphicx}
\usepackage{epsfig}
\usepackage{color}
\definecolor{DarkBlue}{rgb}{0.1,0.1,0.5}
\definecolor{Red}{rgb}{0.9,0.0,0.1}
\definecolor{Green}{rgb}{0.0,0.99,0.0}

\begin{document}

\title{Orbital Order and Spontaneous Orthorhombicity in Iron Pnictides}
\author{C.-C. Chen}
\affiliation{Stanford Institute for Materials and Energy Science, SLAC National Accelerator Laboratory, 2575 Sand Hill Road, Menlo Park, California 94025, USA}
\affiliation{Department of Physics, Stanford University, Stanford, California 94305, USA}
\author{J. Maciejko}
\affiliation{Stanford Institute for Materials and Energy Science, SLAC National Accelerator Laboratory, 2575 Sand Hill Road, Menlo Park, California 94025, USA}
\affiliation{Department of Physics, Stanford University, Stanford, California 94305, USA}
\author{A. P. Sorini}
\affiliation{Stanford Institute for Materials and Energy Science, SLAC National Accelerator Laboratory, 2575 Sand Hill Road, Menlo Park, California 94025, USA}
\author{B. Moritz}
\affiliation{Stanford Institute for Materials and Energy Science, SLAC National Accelerator Laboratory, 2575 Sand Hill Road, Menlo Park, California 94025, USA}
\affiliation{Department of Physics and Astrophysics, University of North Dakota, Grand Forks, North Dakota 58202, USA}
\author{R. R. P. Singh}
\affiliation{Department of Physics, University of California, Davis, California 95616, USA}
\author{T. P. Devereaux}
\affiliation{Stanford Institute for Materials and Energy Science, SLAC National Accelerator Laboratory, 2575 Sand Hill Road, Menlo Park, California 94025, USA}

\date{\today}
\begin{abstract}
A growing list of experiments show orthorhombic electronic anisotropy in the iron pnictides, in some cases at temperatures well above the spin density wave transition. These experiments include neutron scattering, resistivity and magnetoresistance measurements, and a variety of spectroscopies. We explore the idea that these anisotropies stem from a common underlying cause: orbital order manifest in an unequal occupation of $d_{xz}$ and $d_{yz}$ orbitals, arising from the coupled spin-orbital degrees of freedom. We emphasize the distinction between the total orbital occupation (the integrated density of states), where the order parameter may be small, and the orbital polarization near the Fermi level which can be more pronounced. We also discuss light-polarization studies of angle-resolved photoemission, and demonstrate how x-ray absorption linear dichroism may be used as a method to detect an orbital order parameter.
\end{abstract}
\maketitle

The iron pnictides~\cite{discovery} possess a phase diagram similar to the cuprates with superconductivity in close proximity to magnetic order. Upon doping the magnetism is suppressed and superconductivity emerges.~\cite{PD} In contrast, the pnictide parent compounds are metals, rather than insulators, with a Fermi surface consisting of bands mostly from the Fe 3$d$ $t_{2g}$ orbitals. These orbitals form two hole pockets with mostly $d_{xz}$ and $d_{yz}$ characters around the Brillouin zone (BZ) center, and two electron pockets with additional $d_{xy}$ weight at the zone corners.~\cite{ARPES_Lu}

Whether a weak-coupling itinerant electron or a strong-coupling localized moment approach better describes the pnictides remains a topic of debate. It is generally accepted that the Fe on-site, intra-orbital interaction $U$fc~\cite{correlation, seminal-work} is smaller than the overall Fe 3$d$ bandwidth $\sim$ 4eV. In particular, x-ray absorption spectroscopy (XAS) indicates $U\sim$ 2eV. \cite{seminal-work} In this regard, the pnictides are weakly, or at most moderately, correlated materials. On the other hand, both itinerant and local pictures seem to well describe the neutron scattering measurements.~\cite{Dai, neutron_2} The above observations therefore suggest a coexistence of localized moments and itinerant electrons in the iron-pnictide materials.~\cite{coexisting}  They have aspects such as metallicity where correlations play a minor role, while anti-ferromagnetism and local properties derive directly from a strong Hund's coupling.~\cite{seminal-work} This dual character is also implied by other experiments: Raman scattering finds that the $d_{xz}/d_{yz}$ orbitals are more incoherent than $d_{xy}$;~\cite{Raman} de Haas-van Alphen measurements of superconducting LaFePO~\cite{QS_1111} and the closely related non-superconducting compound SrFe$_2$P$_2$~\cite{QS_122} reveal that the electron pockets have a larger mean free path, possibly tied to the $d_{xy}$ band; higher electron mobility is further supported by the Hall effect,~\cite{Hall} as well as nuclear magnetic relaxation.~\cite{NMR}

In addition to this dichotomy, a series of experiments highlight the importance of orthorhombic anisotropy in these materials.~\cite{Dai, JH_1, JH_2, Canfield, STM, NQR} In particular, neutron scattering indicates extremely anisotropic couplings between the Fe moments;~\cite{Dai} scanning tunneling microscopy reveals unidirectional electronic nano-structures,~\cite{STM} and nuclear quadrupole resonance shows a local electronic order in the Fe layers.~\cite{NQR} These observations of broken $C_4$ tetragonal symmetry could stem from an unequal occupation of the $d_{xz}/d_{yz}$ orbitals, arising from the coupled spin-orbital physics.~\cite{OO_Jeroen, OO_Rajiv, spin-orbital, OO_Phillips, OO_Wei, PTM, OO_Frank} A nematic state is formed once the $d_{xz}/d_{yz}$ orbitals are ordered in a translationally invariant way.~\cite{Kivelson} On the other hand, pure spin physics can lead to nematicity~\cite{MM} resulting from order by disorder.~\cite{CCL} The latter scenario can cause an orbital polarization upon breaking $C_4$ symmetry. On symmetry grounds the two scenarios are not distinct since both lead to an Ising order parameter. However, they can be distinguished by the spin-spin correlations and spin dynamics above the spin density wave (SDW) transition.\cite{spin-orbital}

In this study, we assume that orbital occupation is a key driving variable, and investigate the experimental consequences of a net orbital polarization. In particular, we show that recent resistivity anisotropy measurements~\cite{JH_1, JH_2} could result from an unequal $d_{xz}/d_{yz}$ orbital population. Moreover, the Fermi surface reconstruction in angle-resolved photoemission (ARPES)~\cite{LARPES_1, LARPES_2} can be better understood with a (partial) orbital ordering. We also calculate XAS linear dichroism which could be used to probe the orbital order parameter.

In the strong coupling limit, orbital order is easier to understand if the $d_{xz}$ and $d_{yz}$ bands are $\frac{1}{4}$ filled, either with electrons or holes. In this case, the local occupation of the orbitals can be modelled by an Ising variable.~\cite{OO_Rajiv} Such a variable carries a full $R\ln{2}$ entropy,~\cite{spin-orbital} and the orbital order can be robust and easily detectable. In this case, metallic behavior must be due to other orbitals such as $d_{xy}$. The order can survive at intermediate couplings, but is reduced due to $d$-electron  itinerancy.

A more subtle orbital order can arise at intermediate couplings in a system where the occupancy of $d_{xz}$ and $d_{yz}$ orbitals is closer to half filling. In that case, the two orbitals have occupancies of $1+\epsilon$ and $1-\epsilon$. For small $\epsilon$, this can no longer be an atomic scale fluctuating variable but rather one involving a cluster of at least $1/\epsilon$ atoms. Orbital order can arise from an instability that promotes an electron in a local region from the $d_{xz}$ to the $d_{yz}$ orbital (or vice versa). The deviation from unit occupancy is distributed throughout the system. Near half filling, such an electron transfer would raise the system energy due to intra-atomic (Hund's) exchange. A competition therefore results between Hund's energy which favors single occupancy of both orbitals with no orbital order, and inter-atomic exchange which can support orbital order.

If the system is initially frustrated, \emph{e.g.} a $J_1$-$J_2$ model with $J_2$ comparable to $J_1$, this instability to inter-orbital electron transfer can be the system's way of reducing frustration and lowering the inter-atomic exchange energy. This phenomenon also happens in triangular lattice systems which can orbitally order to avoid frustration.~\cite{triangular} In the pnictides, electron transfers can locally break the tetragonal symmetry and couple to lattice distortions. This could make the nearest-neighbor exchange anti-ferromagnetic (AF) in one direction and ferromagnetic (FM) in the other. Thus, collinear AF order will simultaneously minimize all the inter-atomic exchanges. Lowering of this energy would involve every atom in the cluster. Therefore, even if Hund's energy is an order of magnitude larger than the inter-atomic exchange energy, the latter can dominate in substantially large clusters. Apart from anisotropic exchange couplings, an orbital polarization may in turn cause other orthorhombic phenomena, which we discuss below.

In Ba(Fe$_{1-x}$Co$_x$)$_2$As$_2$, a striking in-plane ($ab$) resistivity anisotropy $\Delta\rho/\rho\equiv (\rho_b-\rho_a)/ [\frac{1}{2}(\rho_b+\rho_a)]$, with $\rho_b>\rho_a$, onsets at temperatures well above the structural $T_S$ and magnetic $T_N$ ordering temperatures.~\cite{JH_1, JH_2} Below $T_N$, the magnetic order is AF along the longer $a$-axis and FM along the shorter $b$-axis. One thus expects a higher resistivity along $a$-axis due to reduced hopping. Moreover, if one considers scattering on spin fluctuations,~\cite{Sega, OO_Ashvin, OO_Phillips} stronger scattering along the AF $a$-axis with large $\mathbf{Q}$ momentum transfer is expected; the FM order only gives rise to forward scattering. Both considerations should lead to $\rho_b<\rho_a$, in contradiction to experiment.

It is also puzzling that $\Delta\rho$ increases with decreasing $T$. For scattering on magnetic or lattice fluctuations, $\Delta\rho$ should weaken at lower $T$ because these bosonic degrees of freedom freeze-out. The scattering rate $\Gamma$ due to spin fluctuations is related to the imaginary part of the spin susceptibility, $\Gamma\sim g^2\sum_\mathbf{q}\chi''(\mathbf{q},T)$, where $g$ is the coupling between itinerant electrons and local spins. The divergence of $\chi''$ at the AF wave vector $\mathbf{q}=\mathbf{Q}$ may account for the increased scattering at $T\sim T_N$,~\cite{OO_Phillips} but not for $T$ well above or below $T_N$.

We use a $5$-band model~\cite{Doug} and investigate whether simple impurity scattering, combined with weak orbital polarization, can account for the observed resistivity anisotropy onset at $T>T_S, T_N$. In particular, we show that both the sign and $T$ dependence of $\Delta\rho$ can be qualitatively explained in that way. We study the effect of a small orbital polarization $\langle\delta n\rangle\equiv (n_{xz}-n_{yz})/[\frac{1}{2}(n_{xz}+n_{yz})]$, where $n_{xz}$ ($n_{yz}$) is the $d_{xz}$ ($d_{yz}$) orbital occupation. $\langle\delta n\rangle\neq0$ is mimicked by an effective  on-site energy splitting: $\epsilon_{xz}\equiv \epsilon_0+\Delta_{\textrm{eff}}/2$, and $\epsilon_{yz}\equiv \epsilon_0-\Delta_{\textrm{eff}}/2$. A nonzero $\langle\delta n\rangle$ could result from a spontaneous symmetry breaking mechanism that onsets at $T>T_S,T_N$,~\cite{onset_anomaly} or simply from a small external field $X$ such as applied stress~\cite{JH_2} in the presence of a finite orbital susceptibility $\chi_\mathrm{orb}\propto\partial \langle \delta n \rangle/\partial X$. As $T\rightarrow T_N$, $\langle\delta n\rangle$ and its effect can be enhanced by the coupled spin-orbital physics.~\cite{spin-orbital}

We consider scattering on $\delta$-function nonmagnetic impurities of strength $W$, and calculate the resistivity along the $a,b$-axes using the Kubo formula in the first Born approximation,~\cite{Mahan} where $\Delta\rho$ is weakly dependent on $W$. The potential allows electrons to scatter between different bands but only when the initial and final states have the same orbital character.~\cite{caveat_1} Although thermal fluctuations are not introduced explicitly, their main effect is expected to reduce $\langle\delta n\rangle$ with increasing $T$, i.e. $\Delta_\mathrm{eff}(T)$ should increase as $T$ decreases.~\cite{spin-orbital}

\begin{figure}[t]
\begin{center}
\includegraphics[width=\columnwidth]{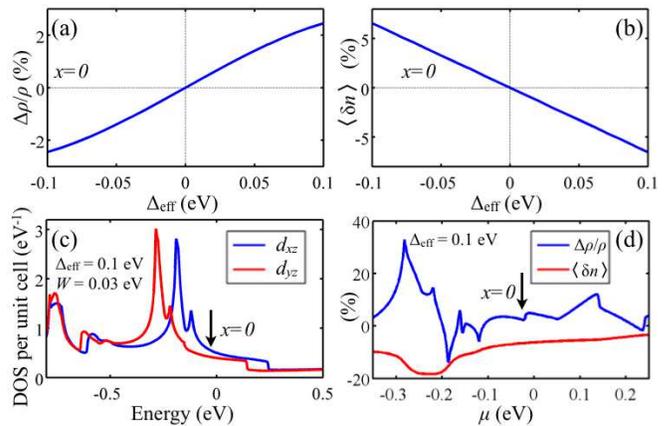}
\end{center} \caption{(Color Online) (a) Resistivity anisotropy $\Delta \rho/\rho$ and (b) orbital polarization $\langle\delta n\rangle$ as a function of the effective energy splitting $\Delta_{\textrm{eff}}$ between the $d_{xz}/d_{yz}$ orbitals calculated from the 5-band model. (c) Anisotropy in the density of states (DOS) with $\Delta_{\textrm{eff}}=0.1$ eV; the arrow indicates the Fermi level at zero doping $x=0$. (d) $\Delta \rho/\rho$ and $\langle\delta n\rangle$ versus the chemical potential $\mu$. Depending on the value of $\mu$, a substantial $\Delta \rho$ could result from a modest $\langle\delta n\rangle$.
}\label{fig:mixed}
\end{figure}

In Fig.~\ref{fig:mixed}(a), a positive splitting $\Delta_\mathrm{eff}>0$ gives rise to the experimentally observed $\rho_b>\rho_a$. Figure~\ref{fig:mixed}(b) displays the corresponding value of $\langle\delta n\rangle$. $\Delta_\mathrm{eff}>0$ corresponds to $\langle\delta n\rangle<0$ (or $n_{xz} < n_{yz}$). On the other hand, \emph{at the chemical potential} ($\mu$) the $d_{xz}$ density of states is larger than the $d_{yz}$ one: $D_{xz}(\mu)>D_{yz}(\mu)$ [Fig.~\ref{fig:mixed}(c)], which is the reason for $\Delta\rho>0$. In the calculation shifting $\mu$ to lower energies would cause sizeable changes in $\delta D(\mu)\equiv D_{xz}(\mu)-D_{yz}(\mu)$ due to the sharp features in the density of states [Fig.~\ref{fig:mixed}(c)], while only changing the overall $\langle\delta n\rangle$ by a small amount. This suggests that a substantial $\Delta \rho$ could result from a modest orbital polarization depending on $\mu$ [Fig.~\ref{fig:mixed}(d)]. Fermi-surface sensitive probes such as transport and ARPES depend strongly on the value of $\delta D(\epsilon)$ near the Fermi level $(E_F)$, which can be appreciable without a large $\langle\delta n\rangle$.

Mean-field calculations of down-folded Hamiltonians~\cite{MFT} find that the collinear AF order is accompanied by a non-zero $\langle \delta n\rangle$. A robust orbital order usually occurs when the $d_{xz}$ $(d_{yz})$ occupation is away from half filling. Moreover, $\langle\delta n\rangle$ depends strongly on interactions. Static Coulomb interaction studies suggest that the electron wavefunctions are ordered along the longer, AF $a$-axis.~\cite{OO_Jeroen, OO_Phillips} On the other hand, LDA calculations based on Wannier orbitals predict an opposite trend.~\cite{OO_Wei} Recent double exchange calculations find that the wavefunction lobes point in the short bond, FM direction due to kinetic energy stabilization.~\cite{OO_Frank} These results indicate that the sign and the magnitude of $\langle \delta n\rangle$ could depend subtly on the interplay between the band-structure and interactions.

In our approach, a nonzero $\Delta\rho$ originates from an orbital-order induced effective mass anisotropy, which if it exists could be detectable by optical conductivity measurements. We note that at low temperatures the resistivity anisotropy can be enhanced due to an anisotropy in carrier density. Recent $ab$-plane optical measurements have reported a depletion of spectral weight upon entering the magnetically ordered state for light polarization \emph{only} along the $b$-axis, with an enhancement of the metallic nature of the charge dynamics related to the $a$-axis response.~\cite{OPC} These observations are indeed consistent with the $dc$ transport measurements. If one also considers scattering on spin fluctuations at $T\sim T_N$,~\cite{OO_Phillips} orbital order can further affect $\Delta\rho$. If $d_{xy}$ electrons are scattered by local moments from the $d_{xz}/d_{yz}$ orbitals, an anisotropy in the spin-fermion coupling $g_{yy}>g_{xx}$ could occur, resulting in $\rho_b>\rho_a$.

\begin{figure}[t!]
\begin{center}
\includegraphics[width=\columnwidth]{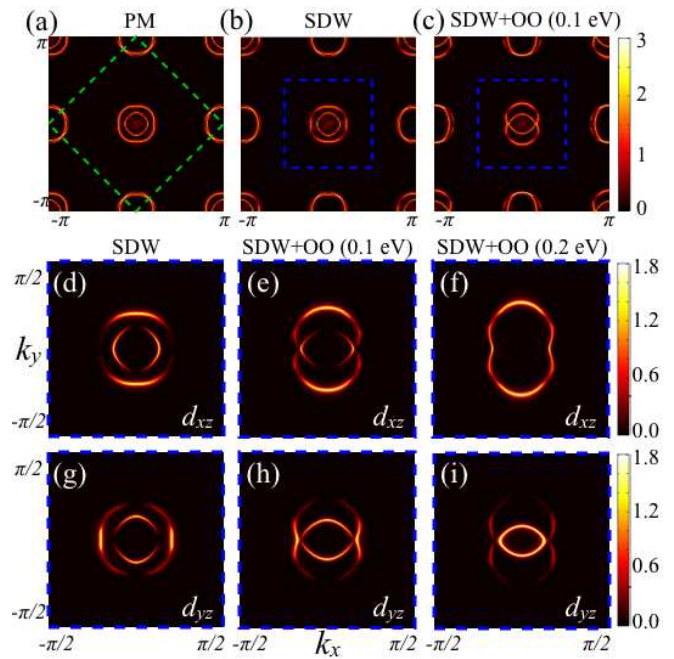}
\end{center}
\caption{(Color Online) (a) $A(\mathbf{k},\omega=E_F)$ in the paramagnetic (PM) phase plotted in the one Fe unit cell BZ; the green box indicates the true two Fe unit cell BZ. (b) $A(\mathbf{k},\omega=E_F)$ in the SDW phase; the blue box indicates the true four Fe unit cell BZ. The shadow bands give rise to only weak spectral weight. (c) $A(\mathbf{k},\omega=E_F)$ in the SDW phase with orbital ordering ($\Delta_{\mathrm{eff}}=0.1$ eV). (d)-(f) The $d_{xz}$ projected spectral functions with SDW and various strength of orbital ordering. (g)-(i) Similar plots for the $d_{yz}$ orbital. When the SDW field is accompanied by a partial orbital ordering, the outer (inner) pocket is dominated by $d_{xz}$ ($d_{yz}$) orbitals. All the plots are broadened with a 10 meV Lorentzian. 
}
\label{fig:ARPES}
\end{figure}

Recent low $T$ ARPES measurements suggest a dominant $d_{xz}$ weight on the hole pockets at the $\Gamma$ point.~\cite{LARPES_1, LARPES_2} Experimentally, the orbital characters can be selected by dipole matrix elements with different light polarizations. Below we use the 5-band model to qualitatively illustrate that a weak SDW field consistent with experiments~\cite{ARPES_Ming} will not result in a significant difference in the $d_{xz}$ and $d_{yz}$ orbital spectral weight. However, a substantial orbital weight reconstruction can be  obtained when the SDW field is accompanied by orbital ordering. The calculations are performed at the mean field level.

Figure \ref{fig:ARPES}(a) shows the spectral function $A(\mathbf{k},\omega=E_F)$ in the paramagnetic phase; the real space unit cell contains two Fe due to the staggered pnictogen height. Figure \ref{fig:ARPES}(b) shows $A(\mathbf{k},\omega=E_F)$  calculated with a symmetry breaking SDW field (=50 meV).~\cite{ARPES_Ming} The real space unit cell now contains 4 Fe, but the shadow bands give rise to only weak spectral intensity. Similarly, $A(\mathbf{k},\omega=E_F)$ in the presence of SDW and a partial orbital order introduced by an effective splitting $\Delta_\mathrm{eff}=0.1$ eV is shown in Fig. \ref{fig:ARPES}(c). Figure \ref{fig:ARPES} (d)-(f) are the $d_{xz}$ projected spectral functions with the SDW field and various strengths of orbital ordering. Figures \ref{fig:ARPES} (g)-(i) are similar plots for $d_{yz}$.

A weak SDW field alone does not result in a significant difference in the spectral weight of the $d_{xz}$ and $d_{yz}$ orbitals. We note that a simple SDW folding cannot account for the ARPES experiments either.~\cite{ARPES_Ming} However, when the SDW field is accompanied by orbital ordering, the outer (inner) pocket is dominated by $d_{xz}$ ($d_{yz}$) character. These results imply an orbital weight reconstruction across the Fermi surface, emphasizing the potential role of the orbital degrees of freedom.

Conclusive evidence for orbital ordering in the iron pnictides could be obtained by performing x-ray dichroism measurements. When the system is structurally distorted or orbitally ordered, the absorption spectrum for incident light polarized along the $x$-axis of the crystal frame would be different from that along the $y$-axis. The resultant difference in the two XAS spectra  $I_x(\omega)-I_y(\omega)$ is referred to as the linear dichroism (LD). A temperature evolution of the LD signals would be highly suggestive of the orbital-ordering hypothesis.

\begin{figure}[t!]
\begin{center}
\includegraphics[width=\columnwidth]{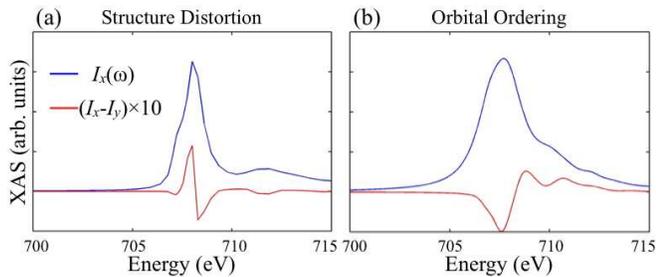}
\caption{(Color Online) (a) The Fe $L_3$-edge XAS for $x$ and $y$ polarization from FEFF calculation for ${\rm BaFe}_2{\rm As}_2$ in the low temperature $T=20K$ orthorhombic phase. The LD signal ($I_x-I_y$) is multiplied by a factor of ten. (b) Atomic multiplet calculation with solely orbital ordering $\Delta_{\textrm{eff}}=0.1$ eV for $L_3$-edge XAS. In both cases, a $\sim 3$\% LD signal results. The structure distortion would saturate at low $T$, but the orbital polarization is expected to increase with decreasing $T$.
}\label{fig:LD} 
\end{center}
\end{figure}

We focus on Fe $L$-edge XAS where the Fe $2p$ core electrons are excited to unoccupied 3$d$ levels. A recent study featured the {\tt FEFF} software~\cite{FEFF_1} to simulate the polarization averaged Fe $L$-edge XAS in a variety of iron-pnictide materials. Results have been obtained in good agreement with experiments.~\cite{seminal-work} Similar calculations including the XAS polarization dependence are shown in Fig. \ref{fig:LD}(a). The LD signal is $\sim$3$\%$ based solely on the orthorhombicity of the low temperature $T=20K$ structure.~\cite{structure}

As the {\tt FEFF} code has no direct control over orbital occupation, we perform an atomic multiplet calculation to simulate the effect of solely orbital ordering. Figure \ref{fig:LD}(b) shows the XAS and a $\sim$3$\%$ LD signal obtained from this method with $\Delta_{\textrm{eff}}=0.1$ eV. The LD magnitude would increase with a stronger $\Delta_{\textrm{eff}}$. We note that the sign and the magnitude of the LD signal and $\langle \delta n \rangle$ are closely related, which may either cancel or enhance the LD signal from the structural distortion.

The LD measurements are complicated by the requirement of single crystals \emph{with de-twinned domains}, which could be achieved by magnetic~\cite{JH_1} or mechanical~\cite{JH_2, Canfield} means. Moreover, below $T_N$ the AF phase also contributes to the LD spectra, so it is important to distinguish these different contributions. A careful temperature dependence should be taken near and across different transition lines. An evolution of the dichroic signal could be observed at high temperatures $T\ge T_S,T_N$ if orbital ordering exists. At low $T$ the structure distortion saturates,~\cite{Berkeley} but the orbital polarization is expected to increase with decreasing $T$.~\cite{spin-orbital} Therefore, these different effects can potentially be distinguished by the LD magnitude and line shape.

In conclusion, we have addressed the in-plane orthorhombic anisotropies in the iron pnictides arising from orbital order. We demonstrated that an unequal $d_{xz}/d_{yz}$ orbital population can lead to a resistivity anisotropy similar to experiments. We also showed that the Fermi surface reconstruction in ARPES is better understood when the SDW field is accompanied by an orbital polarization. We made predictions for XAS linear dichroism which potentially can probe the orbital order parameter. In our view, orbital ordering in these materials arises from coupled spin-orbital fluctuations, where the orbital variable plays a crucial role. Experiments above the SDW transition and theories that address the interplay of spin and orbital fluctuations~\cite{OO_fluctuation} can further clarify the relative importance of the orbital degrees of freedom.

The authors acknowledge discussions with J.-H. Chu, I. R. Fisher, W. Lv, F. Kr\"uger, H.-H. Lai, E. Berg, M. Yi, D. H. Lu, S. Zhou, Z. Wang, L. Degiorgi and P. Thalmeier. This work is supported by the U.S. Department of Energy (DOE) under Contract No. DE-AC02-76SF00515. This research used resources of NERSC, supported by DOE under Contract No. DE-AC02-05CH11231. JM is supported by the Stanford Graduate Program.


\begin{thebibliography}{arprev}
\bibitem{caveat_1}
As the symmetry group of the effective Hamiltonian is $C_4$, for a simple O(3)-invariant potential $V(\mathbf{r})$, all the off-diagonal components between the $t_{2g}$ orbitals such as $\langle d_{xz}|V(\mathbf{r})|d_{yz}\rangle$ vanish. With the rotation invariance of $V(\mathbf{r})$, all the diagonal components in the $t_{2g}$ bases are equal. As the relevant orbitals near $E_F$ are the $t_{2g}$ ones, a diagonal impurity potential is a good approximation.
\end{thebibliography}
\end{document}